\documentclass[usegraphicx]{mn2e}

\title[K2 and MAXI observations of Sco X-1 - Evidence for disc
  precession?]{K2 and MAXI observations of Sco X-1 - Evidence for disc
  precession?}

\author[]
{Pasi Hakala$^{1}$\thanks{E-mail:pahakala@utu.fi}, Gavin Ramsay,$^{2}$, Thomas Barclay$^{3,4}$, Phil Charles$^{5}$\\
$^{1}$Finnish Centre for Astronomy with ESO (FINCA), V\"ais\"al\"antie 20, 
University Of Turku, FIN-21500 Piikki\"o, Finland. \\
$^{2}$Armagh Observatory, College Hill, Armagh, BT61 9DG, UK\\
$^{3}$NASA Ames Research Center, M/S 244-40, Moffett Field, CA 94035, USA\\ 
$^{4}$Bay Area Environmental Research Institute, Inc., 560 Third St. West, 
Sonoma, CA 95476, USA\\
$^{5}$School of Physics and Astronomy, University of Southampton, Highfield, Southampton SO17 1BJ, UK\\
}

\begin{document}

\date{}

\pagerange{\pageref{firstpage}--\pageref{lastpage}} \pubyear{2002}

\maketitle

\label{firstpage}

\begin{abstract}

Sco X-1 is the archetypal low mass X-ray binary (LMXB) and the
brightest persistent extra-solar X-ray source in the sky.  It was
included in the K2 Campaign 2 field and was observed
continuously for 71 days with 1 minute time resolution. In this paper
we report these results and underline the potential of K2 for similar
observations of other accreting compact binaries.  We reconfirm that
Sco X-1 shows a bimodal distribution of optical "high" and "low"
states and rapid transitions between them on timescales less than 3
hours (or 0.15 orbits).  We also find evidence that this behaviour has
a typical systemic timescale of 4.8 days, which we interpret as a
possible disc precession period in the system. Finally, we confirm the
complex optical vs. X-ray correlation/anticorrelation behaviour for
"high" and "low" optical states respectively.
 
\end{abstract}


\section{Introduction}

Sco X-1 is the brightest {\b persistent} extra-solar X-ray source in
the sky (Morrison 1967) and is a low mass X-ray binary (LMXB), where a
Roche-lobe-filling secondary star is losing matter that is accreted by
a neutron star via an accretion disc (Charles \& Coe 2006). There are
numerous studies on this 0.787 d orbital period system over the last
fifty years covering almost the entire electromagnetic spectrum.  It
is also predicted to be a strong source of gravitational waves
(e.g. Aasi et al. 2014).

As the prototype LMXB, Sco X-1 has been studied for decades at all
wavelengths, but especially at optical and X-ray wavelengths.
Ilovaisky et al. (1980) found that the optical and X-ray flux was well
correlated, especially when in a bright state, whilst a longer series
of optical observations showed that Sco X-1 changes from a high to low
state on a timescale of a few hours (Hiltner \& Mook, 1967). McNamara
et al. (2003) presented a year-long optical study of Sco X-1 and
concluded that the optical observations can be accounted for by
variations in the mass accretion rate. Sco X-1 is unique amongst
LMXBs, because to our knowledge, it is the only one showing clearly
bimodal optical states (Hiltner \& Mook, 1967).

The optical light curves of LMXBs are complex and changes are seen on
short time-scales (see for instance Homer et al. 2001 and Hakala et
al. 2009).  The interpretation of these light curves is made more
difficult because of data gaps, the limited duration and cadence of
the observations. Much time and resources have been spent in this
endeavour (e.g. Shih, Charles \& Cornelisse 2011 to name but one).
The possibility of uninterrupted photometric observations of sources by
{\sl Kepler} has therefore led to dramatic discoveries and results in
the field of exo-planets, asteroseismology and accretion
physics. Whilst the original {\sl Kepler} field contained several
dozen accreting cataclysmic variables (e.g. Howell et al 2013), it did
not contain any known LMXB.

Since the loss of two of its reaction wheels, {\sl Kepler} has been
re-purposed into K2, and is now making a series of observations
of fields along the ecliptic plane, where each field is observed for
$\sim$70 days (Howell et al. 2014). The fact that these fields go
through the Galactic plane allows the study of types of objects which
were not present in the {\sl Kepler} field. In the Campaign 2 field,
Sco X-1 was included as one of the target sources. This paper presents
the K2 observations of Sco X-1 and also simultaneous X-ray data using
MAXI. The unprecedented set of optical observations allows us to
characterise the optical behaviour of Sco X-1 in a way not previously
possible.

\section{Observations}

\begin{figure*}
\includegraphics[scale=0.75,angle=90]{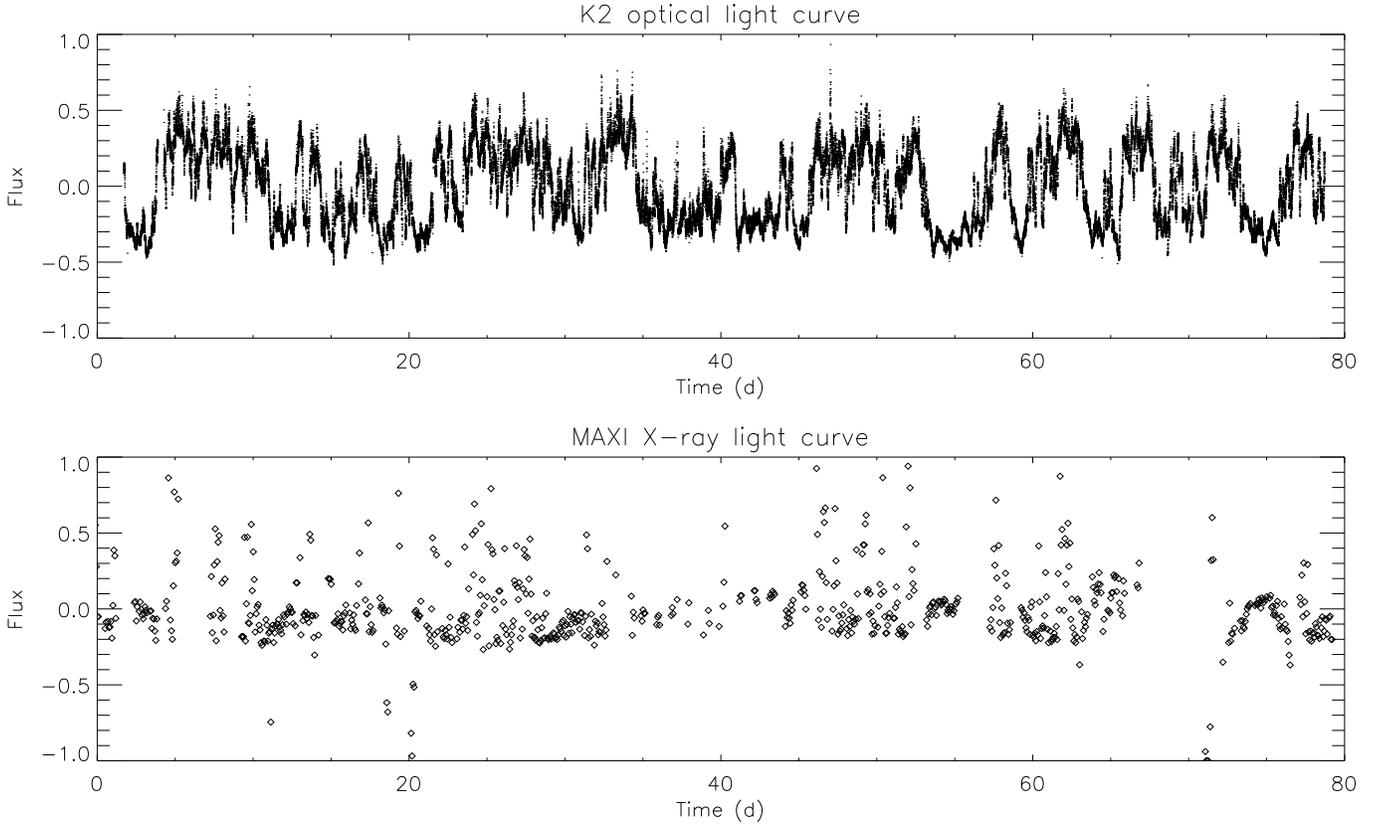}
\caption{The K2 optical (top panel) and MAXI X-ray (lower panel) light
  curves of Sco X-1. The data have been normalized to unity and the
  mean has been subtracted. The error on the optical photometry
    is negligible and the error on the X-ray points are typically 1
    percent.}
\end{figure*}

\subsection{K2 observations}

K2 observations were made of the Campaign field 2 between 2014 Aug 23
and 2014 Nov 10 (MJD = 56892--56971). Both Long Cadence (LC)
data (30 min) and Short Cadence (SC) data (1 min) were obtained from a
14$\times$12 pixel array centered on Sco X-1. Because of a pointing
adjustment early on in the Campaign, data from the first 2.5 days were
omitted, giving a timeline of 71 days. A light curve was made using
{\tt PyKe} software (Still \& Barclay
2012)\footnote{http://keplergo.arc.nasa.gov/PyKE.shtml} which was
developed for the {\sl Kepler} and K2 mission by the Guest Observer
Office. Although the results shown in this work use 1 min time
  resolution data i.e. SC data, we find that the 30 min resolution
  data confirm these results.

Because the {\sl Kepler} satellite lost two of its four reaction
wheels, thrusters are used to periodically re-saturate the reactions
wheels which can then be used to correct for the drift in the
satellite pointing. This results in a significant movement in the
  targets position on the array on a timescale of 6 hr and $\sim$2
  days. Since these systematic effects apply to all of the extracted
  light curves in the K2 field they can be removed using the method
  outlined in Vanderburg \& Johnson (2014). (See Fig 5 of this paper
  to see an example of removing this correlated noise from a light
  curve). The resulting optical (4370-8360\AA) light curve of Sco X-1
(Fig. 1) is unprecedented in the field of optical monitoring of X-ray
binaries. Given the brightness of Sco X-1 ($R\sim$12.4) the
photometric error on each K2 point is negligible.

\begin{figure}
\includegraphics[scale=0.55,angle=0]{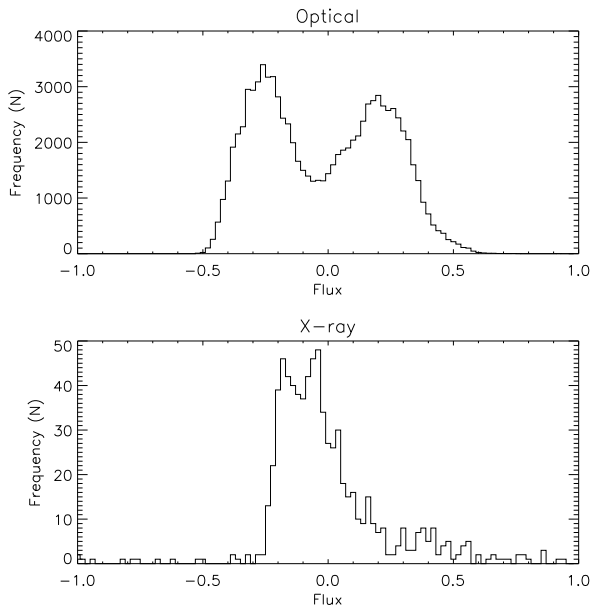}
\caption{The histograms of optical and X-ray fluxes showing the
  bimodal "low"/"high" state behaviour in the optical.}
\end{figure}

\subsection{MAXI observations}

The MAXI all-sky monitor on the International Space Station allows for
the detection and monitoring of bright X-ray sources over the entire
sky in the 2--20 keV band (Matsuoka et al. 2009). Data covering the
time interval of the K2 observations was downloaded from the the MAXI
archive\footnote{http://maxi.riken.jp/top} giving a total of 750
photometric points. Thus, on the average, we obtained one point
approximately every 100 mins (Fig.1). The typical error on an
  individual MAXI point is less than 1\%.

\section{Data Analysis and Discussion}

\subsection{Optical light curve}

The K2 light curve (Fig. 1) immediately reveals that the system has
effectively a bimodal optical brightness distribution, which we refer
to as "low" and "high" states. This is clearly demonstrated by the
light curve, as well as by the histogram of optical fluxes
(Fig. 2). Earlier studies of the optical flux distribution
(eg. McNamara 2003 and the references therein) have yielded more
varied results, which we believe, has been due to the uneven coverage
of the variety of datasets employed. It is worth noting though, that
the 89 h observing campaign of Hiltner \& Mook (1967) produced an
almost identical brightness histogram that presented here (Fig.2).

K2, however, provides us with an unbiased view of the source behaviour
over 71 days. We find that the source spends almost exactly equal
amounts of time in both "low" and "high" states. If we adopt the mean
flux (0.0) as the dividing line between the states, then the system
spends 51 and 49 $\pm$0.3\% in "low" and "high" states respectively.

We have carried out the period analysis of the K2 data using the
Lomb-Scargle power spectrum method (Scargle 1982) since there are some
gaps in the data and the MAXI X-ray data is not equally spaced in
time. The resulting power spectra are plotted in Fig 3. The optical
power spectrum shows a clear signal at 0.787 d, which agrees with the
reported orbital period of the system (Hynes \& Britt, 2012). There
are other peaks in the power spectrum worth commenting on.  Firstly,
there is another peak at about 4.8 d and a third one at around
20d. Since the length of the observation is 71 d, we find that the 20
d peak could easily be produced by red noise effects, even though
there is some tentative evidence in the light curve (Fig 1.) that
extended periods of "high" state might be separated by $\sim$20 days
(see the extended $\sim$ 10d long periods of "high" state beginning
approximately at times 5, 25 and 45d). However, the next extended
"high" state which would have occured at $\sim$ 65d is missing.
  Although the signature of the 4.8 d quasi period is visible in the
  K2 light curve of Sco X-1 (Fig. 1) we have examined and derived
  power spectra of several dozen sources which were observed in the
  same module and chip as Sco X-1 and had a similar brightness. 
  None show any indication of period around 4.8 d. We conclude that the 
  4.8 d quasi-period which we detect in the K2 photometry is intrinsic to Sco X-1.

\begin{figure}
\includegraphics[scale=0.38,angle=90]{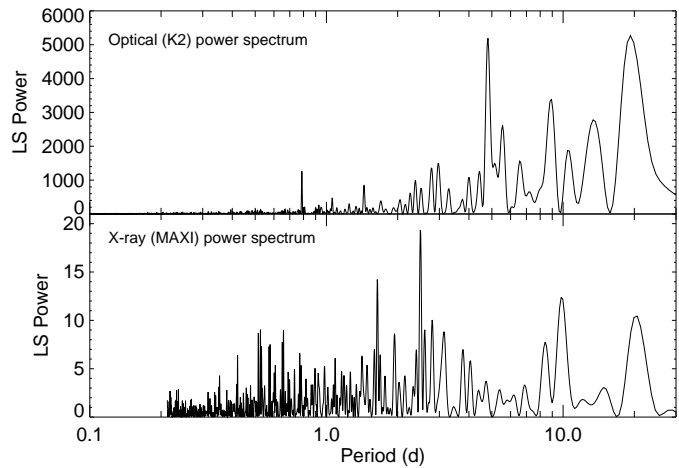}
\caption{The Lomb-Scargle power spectra of the K2 (optical) and
  MAXI (X-ray) light curves.}
\end{figure}

Returning to the 4.8 d period, we can see evidence for transitions
between the "low" and "high" states on this timescale all through the
K2 observation. These transitions are fairly rapid, since they can
take place in less than 3 h, considerably less than Sco X-1's 18.9 h
orbital period. We interpret the 4.8 d peak in the power spectrum as
the main duty cycle for these state changes.  If we fold the light
curve on the orbital period of 0.787 d separately for the "low" and
"high" state data (Fig 4.), it is clear that the shape of the orbital
modulation changes. In "high" state the orbital light curve is almost
sinusoidal, whilst in the "low" state it becomes distinctly
non-sinusoidal in appearance. Furthermore, the amplitude of modulation
is somewhat reduced in the "low" state. This would imply that the
"high" state curve could be produced simply by the X-ray heating of
the inner face of the secondary star. However, the "low" state light
curve requires a variable contribution from an accretion disc, either
by means of phase dependent absorption, emission or changing projected
area of the disc. It is also possible that if the disc is warped out
of the orbital plane, it could partly shield the secondary star from
the X-ray irradiation, thereby diminishing and/or skewing the X-ray
heating effects.

\begin{figure}
\includegraphics[scale=0.5,angle=0]{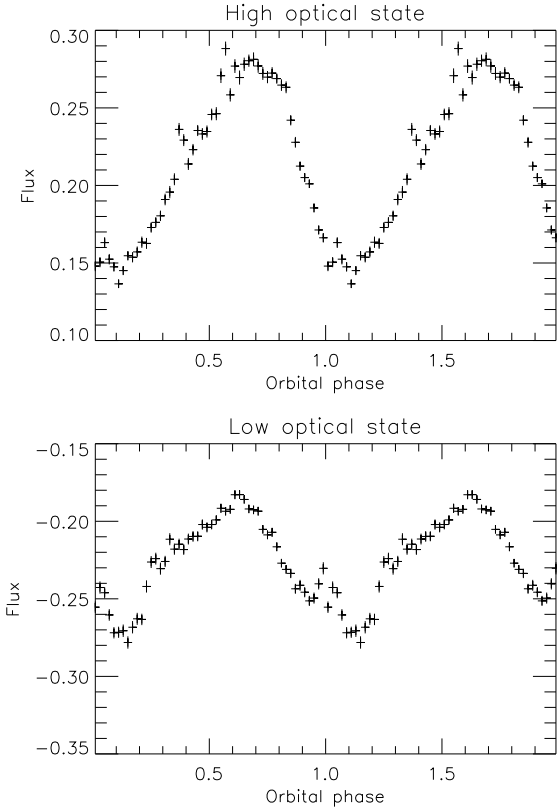}
\caption{The "high" and "low" state data folded separately over the
  orbital period into 50 phase bins and plotted twice for clarity
    (the zero phase is arbitrary). Each phase bin contains
    approximately 900 data points.}
\end{figure}

\subsection{The Optical/X-ray correlation}

McNamara et al. (2003, 2005) demonstrated that the optical and X-ray
emission in Sco X-1 is anti-correlated when the system is in the "low"
state and correlated when in the "high" state. They also showed that
the accretion rate and B magnitude of Sco X-1 are closely related for
most of the optical variability range. Our analysis of K2 and MAXI
data confirms this.  We have binned the optical and X-ray points into
200 common time bins and show our correlation results in Fig 5.
Whilst the correlation behaviour does not appear to be strong, the
rank correlation analysis reveals (carried out separately for the
"low" and "high" state points) that the -0.304 anticorrelation
(Kendall's tau) for the "low" state points has a chance probability of
9.4$\times10^{-6}$. Similarly, for the "high" state points, we obtain
Kendall's tau of 0.305 and a chance probability of
5.4$\times10^{-6}$. We do not see any evidence for the optical and
X-ray fluxes being correlated at the very lowest level of optical
emission as suggested by McNamara et al. (2003).

It is clear from Fig. 1 that the X-ray emission is increased and more
stable during most of the optical minima.  In order to demonstrate
this further, we have plotted the data from the last minimum on a
larger scale in Fig. 6. Evidently the optical data also show much
larger variability than the X-ray data during the minimum. For some
reason, the level of X-ray emission seems to be the same during
several optical minima (Fig. 1) and whenever the X-ray emission reaches
a level higher than this, it becomes more unstable and starts
flaring. It is therefore possible that the X-ray flux during the
optical minima could represent the Eddington limit i.e. when the
system passes from normal branch to the flaring branch in the X-ray
colour-colour Z-diagram (van del Klis 1989). This is, however, not in
agreement with McNamara et al. (2005), where they show that the
accretion rate should be a linear function of B magnitude.

\begin{figure}
\includegraphics[scale=0.55,angle=0]{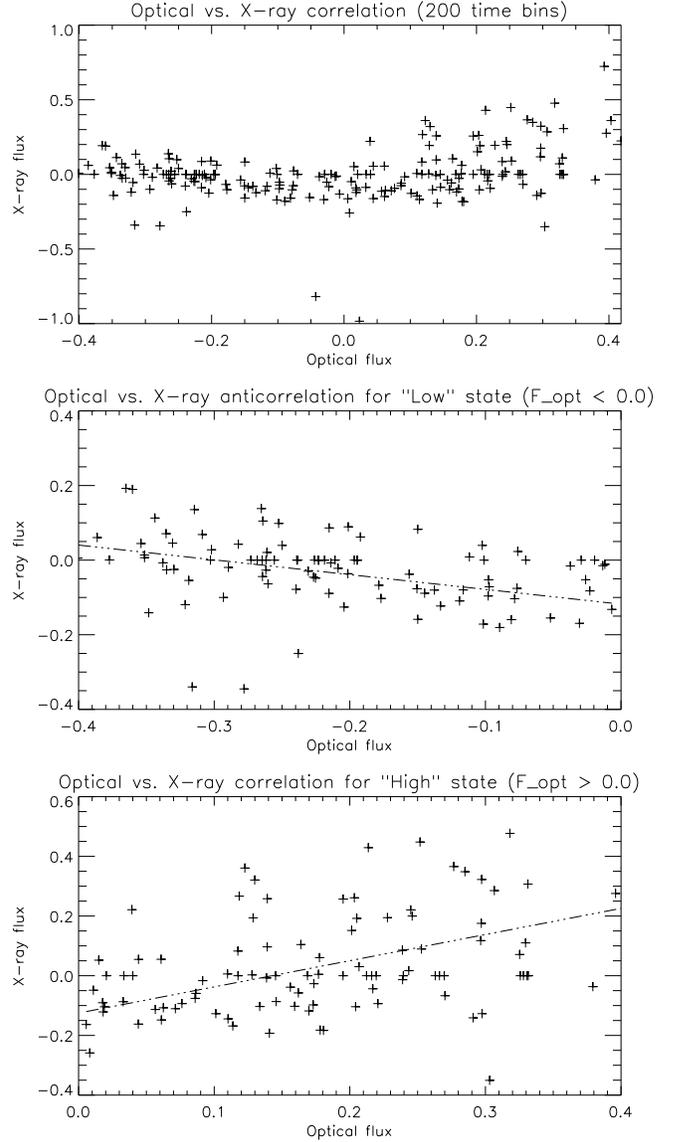}
\caption{The Optical vs. X-ray correlations. All data (top), "low"
  state (middle) and "high" state (bottom). We show the data
    binned into 200 time bins covering the K2 and MAXI observation.}
\end{figure}

\begin{figure}
\includegraphics[scale=0.37,angle=90]{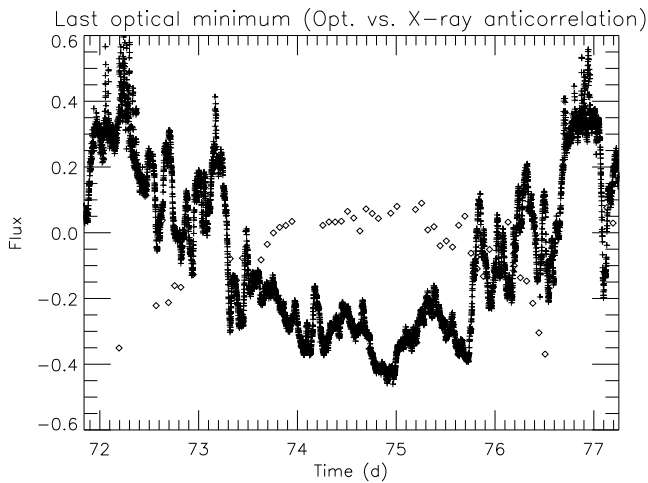}
\caption{We show the last $\sim$5 days of the data shown in Fig. 1
  which includes the last optical minimum and highlights the optical
  X-ray anticorrelation in the "low" optical state.}
\end{figure}

\subsection{The Origin of the 4.8d cycle}

There are several plausible explanations for the origin of the 4.8 d
modulation, some of which we now discuss in detail. It has been shown
(McNamara et al. 2003 and this work) that when the system is in the
"high" state, the X-ray and optical flux correlate. This, together
with the sinusoidal orbital modulation of the optical light curve,
strongly suggests that in the "high" state the disc is flat and
probably does not precess considerably.  This would mean that the
optical orbital modulation is due to the heated face of the secondary
(with a constant emission component from the accretion disc).  This is
supported by the SPH simulations of accretion discs in intermediate
mass ratio $q\sim0.3$ systems (Murray et al. 2000).  The mass ratio in
Sco X-1 is estimated to be $\sim$0.3 (Steeghs \& Casares 2002), which
should make the system stable against the 3:1 resonance when it is
accreting steadily. However, if the accretion rate at L1 drops, it is
plausible that the disc might then start precessing (Murray et
al. 2000).

It has been suggested that the state changes in Sco X-1 are accretion
rate related (Vrtilek et al. 1991, McNamara et al. 2005), in which
case, the appearance of the "low" state folded orbital light curves
could be explained as a result of reduced accretion rate plus
precession (either through changing disc area or changing shadowing of
the secondary from the X-rays). However, this raises two serious
complications.  Firstly, what drives the accretion rate to change on a
time-scale less than 3 hours, as shown by our data?  Furthermore, the
large scale disc structure should not be able to react dramatically to
the accretion rate changes in less than $\sim$1/6 of an orbital
cycle. Secondly, echo mapping of Sco X-1 has revealed that the
accretion disc itself, not the heated face of the secondary, is the
prime source of optical continuum emission, whilst Bowen blend
emission lines are reprocessed on the secondary (Mu\~noz-Darias et
al. 2007). In general, the optical and UV emission lines can be
produced both in the disc and in the heated face of the secondary
(Steeghs \& Casares 2002, Boroson, Vrtilek \& Raymond 2014). The
cross-correlation analysis of HST UV data with RXTE data (Kallman et
al. 1998) produced somewhat mixed results with the UV continuum and
lines displaying inverse behaviour in relation to the X-ray data.

The second possibility for the 4.8 d period is disc precession.
Assuming that 4.8 d is indeed the disc precession period, and given
the known 0.787 d orbital period, then this would imply a
beat/superhump period of 0.94 d, assuming prograde precession of the
disc. This would further imply a superhump period excess
$\epsilon$=0.20 which, although rather large, is still broadly
compatible with that obtained from the SPH simulations for $q$=0.3
($\epsilon$=0.16, Murray et al. 2000).  However, there is no clear
sign of the expected 0.94 d period in the power spectrum. One
explanation for this could be that the superhump period is thought to
arise in CVs as a result of the changing free fall length/potential
well depth of the stream and disc impact point (Whitehurst \& King
1991). Whilst this is the case in CVs where the disc is viscously
heated, the impact point might not be significantly hotter than the
surrounding disc in a system like Sco X-1, where the disc is
predominantly X-ray heated. If the 4.8 d period is indeed due to the
disc precession, it is still not clear how this could drive the
"low"/"high" state behaviour with such short transition times (less
than 3h) from one state to another.

\section{Conclusions}

The K2 observations of Sco X-1 have given the X-ray binary
  community a virtually uninterrupted and unprecedented optical light
  curve of a LMXB covering more than 70 days. It shows the detailed
  investigation of the complex relationship between the optical and
  X-ray flux in a way not previously possible. Furthermore, K2 data
has clearly demonstrated that Sco X-1 has bimodal optical states with
rapid ($<$3 h) transitions from state to state.  There is also
evidence that these transitions possibly occur periodically with a
period of 4.8 d. As earlier studies have linked the optical brightness
strongly with the accretion rate (McNamara et al. 2005), we conclude
that the accretion rate could vary with a period of 4.8 d. This, in
turn, could be due to the precession of the accretion disc at such a
period. Further observations of the accretion disc structure
(eg. Doppler mapping) over this 4.8 d cycle are encouraged to verify
the possible periodic changes in the disc geometry.

\section{Acknowledgments}

Funding for the K2 spacecraft is provided by the NASA Science Mission
Directorate. The data presented in this paper were obtained from the
Mikulski Archive for Space Telescopes (MAST). This research has made
use of the MAXI data provided by RIKEN, JAXA and the MAXI team. Our
work has made use of PyKE, a software package for the reduction and
analysis of Kepler and K2 data. This open source software project is
developed and distributed by the NASA Kepler Guest Observer
Office. Armagh Observatory is supported by the Northern Ireland
Government through the Dept Culture, Arts and Leisure. 
After the submission of this Letter, a paper by Scaringi et al. (2015) appeared 
in astro-PH. Apart from the 4.8d period (which they do not refer to), their analysis of theÊ
same dataset is broadly in agreement with our results.

\end{document}